\documentclass{article}
\usepackage{amsmath}
\usepackage{amssymb}
\usepackage{booktabs}
\usepackage{amsmath}
\usepackage{amsfonts}
\usepackage[backend=bibtex,url=false,doi=false,style=numeric-comp,firstinits=true]{biblatex}
\renewbibmacro{in:}{}
\bibliography{/home/rsrsl/library/jabref_biblio.bib}
\usepackage{enumerate}
\usepackage{graphicx}
\graphicspath{{images/}}
\usepackage{float}
\usepackage{algorithm}
\usepackage{xcolor}
\usepackage{epstopdf}
\newcommand{\norm}[1]{\left\| #1 \right\|}

\newcommand{\E}{\mathbb{E}}
\newcommand{\Var}[1]{\mathrm{Var}\left\lbrace#1\right\rbrace}
\newcommand{\trace}{\mathrm{Tr}}
\newcommand{\T}{\mathbf{T}}
\newcommand{\R}{\mathbf{R}}
\newcommand{\kron}{\otimes}
\newcommand{\N}{\ensuremath{\mathcal{N}}}

\newcommand{\matlab}{\textsc{matlab}}

\renewcommand{\Vec}{\mathrm{vec}}
\newcommand{\vect}{\mathrm{vec}}
\renewcommand{\matrix}[1]{\begin{bmatrix} #1 \end{bmatrix}}

  \newtheorem{rem}{Remark}
  \newtheorem{thm}{Theorem}

\title{A kernel-based approach to Hammerstein system identification} 
\author{Riccardo Sven Risuleo, Giulio Bottegal, and H\r akan Hjalmarsson
\thanks{Riccardo S. Risuleo, Giulio Bottegal, and H\r akan Hjalmarsson are with the
ACCESS Linnaeus Centre, School of Electrical Engineering,
KTH Royal Institute of Technology, Stockholm, Sweden (e-mail: \{risuleo;
bottegal; hjalmars\}@kth.se). This work was supported by the European Research Council
  under the advanced grant LEARN, contract 267381 and by the Swedish Research
Council under contract 621-2009-4017}} 
\begin{document}
\maketitle

\begin{abstract}                          
In this paper, we propose a novel algorithm for the identification of
Hammerstein systems. Adopting a Bayesian approach, we model the impulse
response of the unknown linear dynamic system as a realization of a zero-mean
Gaussian process. The covariance matrix (or kernel) of this process is given by
the recently introduced stable-spline kernel, which encodes information on the
stability and regularity of the impulse response. The static nonlinearity of
the model is identified using an Empirical Bayes approach---that is, by maximizing
the output marginal likelihood, which is obtained by integrating out the
unknown impulse response. The related optimization problem is solved adopting a
novel iterative scheme based on the Expectation-Maximization method, where
each iteration consists in a simple sequence of update rules. Numerical
experiments show that the proposed method compares favorably with a standard
algorithm for Hammerstein system identification.
\end{abstract}


\section{Introduction}
The Hammerstein system is a cascaded system composed of a static nonlinearity
followed by a linear dynamic system (see e.g.~\cite{ljung1999system}).  Hammerstein system
identification has become object of research apparently since the Sixties
(see~\cite{narendra1966iterative}). Due to the wide spectrum of applications (see
e.g.~\cite{hunter1986identification},~\cite{westwick2001separable},
\cite{bai2009identification}), Hammerstein system identification has gained
popularity through the years and a wide range of methods has been developed
(see~\cite{rangan1995new},~\cite{bai1998optimal},~\cite{bai2004convergence}, and
references therein).

Several identification approaches have been proposed. For
instance,~\cite{greblicki1986identification} exploits kernel regression
arguments,~\cite{westwick2001separable} uses a separable least squares
approach,~\cite{greblicki2002stochastic} focuses on stochastic system identification of
Hammerstein models, while~\cite{goethals2005subspace} proposes a subspace
approach. Research on this topic is still rather active
(see~\cite{schoukens2011parametric},~\cite{han2012hammerstein}).

In this paper, we propose a novel method for Hammerstein system identification.
Following recent developments in identification of linear dynamic systems
(see~\cite{chen2012estimation},~\cite{pillonetto2014kernel}), we adopt a kernel-based
identification approach for the linear part of the Hammerstein model. To this
end, we model the impulse response of the unknown linear system as a
realization of a Gaussian process. The covariance matrix (or kernel) of this
process has a specific structure given by the  \emph{stable spline kernel}
(see~\cite{pillonetto2010new},~\cite{pillonetto2011kernel},~\cite{bottegal2013regularized}). This structure
induces properties such as BIBO stability and smoothness in the Gaussian
process realizations and depends  on a \emph{shaping parameter} which regulates
the exponential decay of the generated impulse responses.

In the context of Hammerstein system identification, we can define an effective
estimator of the linear dynamic block using Bayesian arguments by exploiting
the kernel-based framework. Such an estimator is function of the static
nonlinearity, as well as the kernel shaping parameter and the noise variance. A
crucial point of the proposed approach is the estimation of these quantities.
Exploiting a Bayesian interpretation of kernel-based methods
~\cite{maritz1989empirical}, we perform this estimation step by maximizing
the marginal likelihood of the output measurements, which is obtained by
integrating out the unknown impulse response. This approach has been shown to
be effective in kernel-based linear system identification
~\cite{pillonetto2014tuning}. However, when applied to Hammerstein system
identification, the related optimization problem becomes more involved due to the
presence of the unknown static nonlinearity. To overcome this difficulty, we
propose a novel iterative solution scheme based on the Expectation-Maximization
method proposed by~\cite{dempster1977maximum}. We show that the resulting Hammerstein system
identification algorithm has a rather low computational burden. Remarkably, the
proposed method does not need any parameter to be set nor requires the user to
select the model order of the linear system. This in contrast with standard
parametric methods, where, when little is known about the system, one as to
estimate the optimal model order using complexity criteria or cross validation
(see e.g.~\cite{ljung1999system}).

The method used in this paper is also used in~\cite{bottegal2015blind} in the
context of blind system-identification.

The structure of the paper is as follows. In the next section, we formulate the
Hammerstein system identification problem. In Section~\ref{sec:model}, we
describe the model adopted for the linear system and the static nonlinearity.
In Section~\ref{sec:marginal}, we introduce the proposed algorithm, which is
tested in Section~\ref{sec:experiments}. Some conclusions end the paper.

\section{Problem formulation}
\begin{figure}[H]
  \centering
  \includegraphics[width=0.7\columnwidth]{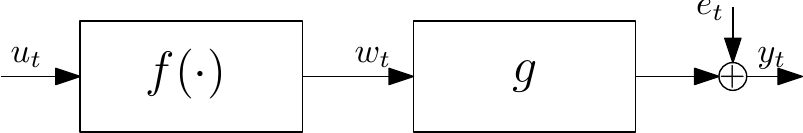}
  \caption{Block scheme of the Hammerstein system.}\label{fig:hammerstein}
\end{figure}

We consider a single input single output discrete-time system described by the
following time-domain relations (see Figure~\ref{fig:hammerstein})
\begin{equation} \label{eq:hammerstein_basic}
\begin{array}{lcl}
	w_t &=& f(u_t) \\
  y_t &=& \sum_{k=1}^{\infty}g_k w_{t-k} + e_t \,.
\end{array}
\end{equation}
In the above equation, $f(\cdot)$ represents a (static) nonlinear function
transforming the measurable input $u_t$ into the unavailable signal $w_t$, which
in turn feeds a linear time-invariant (LTI) strictly causal system described by
the impulse response $g_t$. The output measurements of the system $y_t$ are
corrupted by white Gaussian noise, denoted by $e_t$, which has unknown variance
$\sigma^2$. For simplicity, we assume null initial conditions.
We assume that $N$ input-output samples are collected, and denote them by
${\{u_t\}}_{t=0}^{N-1}$, $i{\{y_t\}}_{t=1}^{N}$.  For notational convenience, we also
assume null initial conditions.  Then, the system identification problem we
discuss in this paper is the problem of estimating $n$ samples of the impulse
response say, ${\{\hat g_t\}}_{t=1}^n$ (where $n$ is large enough to capture the
system dynamics), as well as the static nonlinearity $f(\cdot)$.

\begin{rem}
The identification method we propose in this paper is not affected by the choice
of $n$. Furthermore, it can be derived also in the continuous-time setting,
using the same arguments as in~\cite{pillonetto2010new}. However, for ease of exposition,
here we focus only on the discrete-time case.
\end{rem}

\subsection{Non-uniqueness of the identified system}\label{sec:identifiability}
It is well-known (see e.g.~\cite{bai2004convergence}) that the two components of
a Hammerstein system can be determined up to a scaling factor. In fact, for
every $\alpha \in \mathbb R$, the pair $(\alpha
g_t,\,\frac{1}{\alpha}f(\cdot))$, describes the input-output relation equally
well. We will address this issue in the modeling of the impulse response by
fixing the kernel scaling parameter (see Subsection~\ref{ss:kernel_lti}).

\section{Modeling and identification of Hammerstein Systems}\label{sec:model}
In this section, we first introduce the models adopted for the input static
nonlinearity and the LTI system. Then, we describe the proposed system
identification method.
\subsection{Notation}
Let us introduce the vector-based notation
\begin{equation*}
u \triangleq \begin{bmatrix} u_0 \\ \vdots \\ u_{N-1} \end{bmatrix} ,\,
y \triangleq \begin{bmatrix} y_1 \\ \vdots \\ y_N \end{bmatrix} ,\,
g \triangleq \begin{bmatrix} g_1 \\ \vdots \\ g_n \end{bmatrix} ,\,
e \triangleq \begin{bmatrix} e_1 \\ \vdots \\ e_N \end{bmatrix} .
\end{equation*}
Furthermore, we define the operator $\T_n(\cdot)$ that, given a vector of
length $N$, maps it to an $N\times n$ Toeplitz matrix:
\begin{equation*}
 \T_n(w) =
 \begin{bmatrix} w_0 & 0  & & \cdots & 0 \\
                 w_1 & w_0 & 0 & \cdots & 0 \\
                 \vdots & \vdots &  &\ddots & \vdots \\
                 w_{N-2} & w_{N-3}  &  \cdots & w_{N-n+1}& 0 \\
                 w_{N-1} & w_{N-2}  &  \cdots &\cdots &  w_{N-n}
  \end{bmatrix} \, \in \, \mathbb{R}^{N \times n}  \,.
\end{equation*}
We shall reserve the symbol $W$ for $\T_n(w)$. This allows us to write the input-output relation
of the LTI system as
\begin{equation}\label{eq:linreg_z}
y = Wg + e  \,.
\end{equation}

\subsection{The input static nonlinearity}
Following~\cite{bai1998optimal},~\cite{bai2004convergence}, we assume that
the input static nonlinearity belongs to a $p$-dimensional space of functions
and thus can be described using a linear combination of known basis functions
${\{\phi_i\}}_{i=1}^p$.
Hence, we can write
\begin{equation}
  w_t = f(u_t) = \sum_{i=1}^p c_i \phi_i(u_t) \,,
\end{equation}
where the coefficients $c_i$ are unknown. The problem of estimating $f(\cdot)$
is thus equivalent to the problem of determining such coefficients.
By introducing the following matrix
\begin{equation}
  F(u) \triangleq
  \matrix{\phi_1(u_0)  & \cdots & \phi_p(u_0)\\
             \vdots& \vdots & \vdots\\
             \phi_1(u_{N-1})  & \cdots & \phi_p(u_{N-1})\\
       }
\end{equation}
we can write
\begin{equation}\label{eq:effona}
w = F(u) c \,,
\end{equation}
where $c \triangleq \matrix{ c_1 & \cdots & c_p}^T$.

\subsection{Kernel-based modeling of the LTI system}\label{ss:kernel_lti}
In this paper, we adopt the kernel-based identification approach for LTI
systems, proposed in~\cite{pillonetto2010new},~\cite{pillonetto2011kernel}. To this end, following a
Gaussian process regression approach~\cite{rasmussen2006gaussian}, we assume that the
impulse response of the system is a realization of a zero-mean Gaussian process,
namely
\begin{equation}\label{eq:prior_g}
    g \sim \N(0,\lambda K_{\beta}) \,.
\end{equation}
The matrix $K_\beta$, which is also known as {\it kernel}, is a covariance
matrix parameterized by a shaping parameter $\beta$, and $\lambda \geq 0$ is a
scaling factor. In the context of system identification, the family of the
\emph{stable spline
kernels}~\cite{pillonetto2010new},~\cite{pillonetto2011kernel}itutes a valid
choice, since they promote BIBO stable and smooth realizations. Specifically, we
employ the \emph{first-order stable spline kernel} (or \emph{TC kernel}
in~\cite{chen2012estimation}) given by
\begin{equation}\label{eq:ssk1}
  {\{K_\beta\}}_{i,j} \triangleq \beta^{ \max(i,j)} \,,\quad  0\leq\beta<1
\end{equation}
where $\beta$ determines the decaying velocity of the generated impulse
responses. As $\lambda$ regulates the amplitude of the impulse response $g$, we
can we can arbitrarily set $\lambda = 1$, to cope with the identifiability issue
described in Section~\ref{sec:identifiability}.

\subsection{Estimation of the LTI system}

In this section we derive the system identification strategy that arises
naturally when kernel-based methods are employed. The estimator we will obtain
is function of the vector
\begin{equation}\label{eq:hyperparameters}
\theta \triangleq \begin{bmatrix} c^T  & \sigma^2 & \beta \end{bmatrix} \quad \in \mathbb{R}^{p+2} \,,
\end{equation}
which we shall call \emph{hyperparameter vector}. Since the noise is assumed
Gaussian, the joint distribution of the vectors $y$ and $g$ is again Gaussian,
and parametrized by $\theta$.
Thus
\begin{equation}\label{eq:joint_Gaussian}
p\left(\begin{bmatrix} y \\ g \end{bmatrix};\theta \right) \sim \mathcal N \left( \begin{bmatrix} 0\\0 \end{bmatrix} , \begin{bmatrix} \Sigma_y & \Sigma_{yg} \\ \Sigma_{gy} &  K_\beta \end{bmatrix} \right)\,,
\end{equation}
where $\Sigma_{yg} = \Sigma_{gy}^T =   W K_\beta$ and $\Sigma_y =  W K_\beta W^T + \sigma^2I$.
It follows that the posterior distribution of $g$ given $y$ is also
Gaussian:
\begin{equation}\label{eq:pg}
p(g|y;\,\theta) = \mathcal N \left(Cy,\,P \right) \,,
\end{equation}
where
\begin{equation}\label{eq:CandP}
  P = {\left( \frac{W^T W}{\sigma^2} +  K_\beta^{-1} \right)}^{-1} \quad,\quad C = P \frac{W^T}{\sigma^2}  \,.
\end{equation}
The hyperparameter vector $\theta$ needs to be estimated from the
available data, and we will address this point in the next section.

Given a value of $\theta$, from~\eqref{eq:pg}, we find the impulse response
estimator as the minimum mean squared error
estimator~\cite{anderson2012optimal}
\begin{equation}\label{eq:Bayesest_gaussian}
\hat g = \mathbb E [g|y;\,\theta] = C y \,.
\end{equation}

\section{Empirical Bayes estimates of the parameters}\label{sec:marginal}
In this section we deal with the estimation of the hyperparameter vector
$\theta$. Exploiting the Bayesian framework introduced in the previous section,
we adopt an Empirical Bayes approach~\cite{maritz1989empirical} for this task. The
hyperparameter vector is obtained by maximizing the marginal likelihood of the
output:
\begin{equation}\label{eq:marg_lik}
  \begin{split}
  \hat \theta &= \arg\max_{\theta} \log p(y;\theta) \\
							&= \arg\min_{\theta} \log \det \Sigma_y + y^T
                            \Sigma_y^{-1} y.
  \end{split}
\end{equation}

\subsection{Solution via the EM method}
Problem~\eqref{eq:marg_lik} is non-convex and nonlinear, and involves $p+2$
decision variables. For its solution, we propose a scheme based on the EM
method. To this end, we introduce the complete-data log-likelihood
\begin{equation}\label{eq:complete_likelihood}
L(y,\,g;\theta)  \triangleq \log p(y,\,g;\theta) \,,
\end{equation}
where $g$ plays the role of \emph{latent variable}.
The EM method solves~\eqref{eq:marg_lik} by iteratively marginalizing out $g$
from~\eqref{eq:complete_likelihood}. This operation is performed by iterating the following steps:
\begin{description}
    \item[(E-step)] At the $k$-th iteration, using the estimate
      $\hat \theta^{(k)}$, compute
        \begin{equation}
      \mathcal{Q}(\theta,\,\hat\theta^{(k)}) \triangleq \E_{p(g;y,\,\hat\theta^{(k)})}\left[L(y,\,g;\theta) \right] \,;
    \end{equation}
  \item[(M-step)] update the the estimate solving
    \begin{equation}
      \hat \theta^{(k+1)} = \arg\max_\theta \mathcal{Q}(\theta,\,\hat\theta^{(k+1)}) \,.
    \end{equation}
\end{description}
Such a procedure converges to a maximum (not necessarily global)
of~\eqref{eq:marg_lik} (see e.g.~\cite{mclachlan2007maximum}).

Let us assume that the estimate $\hat\theta^{(k)}$ of the hyperparameter vector
has been computed at the $k$-th iteration of the EM method.
Using~\eqref{eq:CandP}, we can compute the quantities $\hat P^{(k)}$ and
$\hat m_g^{(k)}$, namely the posterior mean and variance of $g$ given $y$. The
following theorem illustrates how to compute $\hat\theta^{(k+1)}$.

\begin{thm}\label{th:main}
Assume that $\hat\theta^{(k)}$ is available. Then, the updated estimate
\begin{equation}
\hat\theta^{(k+1)} = \matrix{\hat c^{(k+1)T}  & \hat \sigma^{2,(k+1)} & \hat \beta^{(k+1)}}
\end{equation}
is obtained by means of the following steps:
\begin{itemize}
  \item The  coefficients of the nonlinear block are given by
    \begin{equation}\label{eq:new_c}
      \hat c^{(k+1)} = {(\hat A^{(k)})}^{-1} \hat b^{(k)} \,,
    \end{equation}
    where
   \begin{equation}\label{eq:A_b}
      \begin{split}
        \hat A^{(k)} &= {F(u)}^T\R^T \big( (\hat P^{(k)} + \hat m_g^{(k)}
        \hat m_g^{(k)\,T}) \kron I_N  \big)\R F(u) \,,\\
        \hat b^{(k)} &= {F(u)}^T \T_N\big(\hat m_g^{(k)}\big)y \,,
      \end{split}
    \end{equation}
    where $\R\in \mathbb{R}^{Nn\times N}$ is the (unique) matrix such that, for
    all $u\in \mathbb{R}^{N}$, we have
    \begin{equation}\label{eq:devec}
      \R u = \Vec\big(\T_n(u)\big);
    \end{equation}
		\item The noise variance is updated using
    \begin{equation}\label{eq:new_sigma}
      \begin{split}
			\hat \sigma^{2,(k+1)} & = \frac{1}{N}\Big(\Vert y - \hat W^{(k+1)}\hat m_g^{(k)}\Vert^2_2 \\
      & +\trace\{\hat W^{(k+1)}\hat P^{(k)}\hat W^{(k+1)T}\}\Big)
			\end{split}
		\end{equation}
    where $\hat W^{(k+1)} = \T_n\big(F(u)\hat c^{(k+1)}\big)$ results by
    plugging the new estimates $\hat c^{(k+1)}$ in~\eqref{eq:effona};
  \item The updated kernel shaping parameter is solution of
    \begin{equation}\label{eq:new_beta}
    \hat\beta^{(k+1)} = \arg \max_{\beta} Q_\beta(\beta,\hat \theta^{(k)})  \,,
    \end{equation}
    where
    \begin{equation}\label{eq:Q_beta}
    Q_\beta(\beta, \hat \theta^{(k)}) \triangleq \log\det K_\beta + \trace\Big[
    K_\beta^{-1}\big(\hat P^{(k)} + \hat m_g^{(k)}\hat m_g^{(k)T}\big)\Big].
    \end{equation}
    \end{itemize}
\end{thm}
Therefore, the solution of~\eqref{eq:marg_lik} can be retrieved in a simple and
quick way. In fact, Theorem~\ref{th:main} states that, given an estimate of the
hyperparameter vector, the updated values of the coefficients of the nonlinear
block are obtained solving a system of linear equations. Then, the new estimate
of the noise variance can also be computed using a closed-form expression.
Finally, the new value of the kernel shaping parameter is retrieved by solving a
simple optimization problem. Although such a problem does not seem to admit a
closed-form solution, we note that it is a one dimensional problem in the
domain $[0,\,1]$. Hence, it can be quickly solved by pointwise evaluation.

Below, we give our novel kernel based method for the identification of
Hammerstein systems.
\begin{algorithm}[ht!]\label{alg}
\textbf{Algorithm}: Kernel-based Hammerstein System Identification  \vspace{0.1cm}\\
Input: ${\{u_t\}}_{t=0}^{N-1}$, ${\{y_t\}}_{t=1}^N$ \vspace{0.1cm}\\
Output: ${\{\hat{g}\}}_{t=1}^n$, $\hat{f}(\cdot)$
\begin{enumerate}
  \item Initialization: randomly set $\hat \theta^{(0)}$
\item Repeat until convergence:
  \begin{enumerate}
    \item  \textbf{E-step:} update $\hat P^{(k)}$, $\hat C^{(k)}$ from~\eqref{eq:CandP}
      and $\hat m_g^{(k)}$ from~\eqref{eq:Bayesest_gaussian};
    \item  \textbf{M-step:} update the parameters:
      \begin{itemize}
        \item $\hat c^{(k+1)}$ from~\eqref{eq:new_c};
        \item $\hat \sigma^{(k+1)}$ from~\eqref{eq:new_sigma},
        \item $\hat \beta^{(k+1)}$ from~\eqref{eq:new_beta}
      \end{itemize}
  \end{enumerate}
\item Compute ${\{\hat{g}\}}_{t=1}^n$ from~\eqref{eq:Bayesest_gaussian}
  and $\hat f(\cdot) = \sum_{i=1}^p \hat c_i \phi_i(\cdot)$;
\end{enumerate}
\end{algorithm}
\begin{rem}
The choice of random initial is motivated by the fact that, after
several numerical experiments we have noticed that the algorithm is capable of
reaching the global maximum of~\eqref{eq:marg_lik} independently of the initial
conditions.
\end{rem}

\section{Numerical Experiments}\label{sec:experiments}
In order to assess the performance of the proposed Hammerstein system
identification scheme, we run a set of Monte Carlo experiments. Specifically, we
perform 8 different identification experiments, each consisting of 100
independent Monte Carlo runs. Depending on the experiment, we generate systems
of order $\nu$, where $\nu = 4,\,8,\,10,\,20$. At each Monte Carlo run, a system
is generated by picking $\nu$ random poles and zeros. The poles and zeros are
located in the set $\{z \in \mathbb{C} \mbox{ s.t. } 0.4 \leq z \leq 0.93\}$.
The input nonlinearity is chosen to be  a polynomial of sixth order, so that
$\phi_i(u) = u^{i-1}$, $i = 1,\,\ldots,\,7$. The roots of the polynomial are
in random locations within the interval $[-2,2]$. The input to the system is
white noise, with uniform distribution in the same interval.  We generate
$N=500$ input/output samples for any Monte Carlo run. The output is corrupted by
Gaussian white noise whose variance is chosen so to obtain a certain signal to
noise ratio, according to
\begin{equation}
  \sigma^2 = \frac{\Var{Wg}}{\mathrm{SNR}} \,,
\end{equation}
where $\mathrm{SNR}$ is either 10 or 1, depending on the experiment. The
features of the 8 experiments are summarized in Table 1.
\begin{table}[h!]
\begin{center}
\begin{tabular}{ccccccccc}
  \toprule
Experiment $\#$ & 1 & 2 & 3 & 4 & 5 & 6 & 7 & 8 \\
\midrule
LTI System Order & 4 & 8 & 10 & 20 & 4 & 8 & 10 & 20 \\
SNR & 10 & 10 & 10 & 10 & 1 & 1 & 1 & 1 \\
\bottomrule
\end{tabular}
\vspace{2mm}
\caption{Features of the 8 Monte Carlo experiments performed to test the proposed method.}\label{tab:fits}
\end{center}
\end{table}
The goal of the experiments is to estimate $n=100$ samples of the impulse
response of the LTI system and the $p=7$ coefficients of the nonlinear block.
In order to obtain uniqueness in the decompositions, we impose $\|g\|_2=1$, and
we assume the sign of the first nonzero element of $g$ to be known.

We compare the following two estimators:
\begin{itemize}
  \item \textbf{KB-H} This is the proposed kernel-based Hammerstein system identification method, which
    estimates the prior shaping parameter $\beta$, the nonlinear
    function and the noise variance through marginal likelihood maximization and the EM method. The convergence criterion for the EM method is $\|\hat{\theta}^{(k+1)} - \hat{\theta}^{(k)} \|_2 < 10^{-3}$.
  \item \textbf{NLHW} This is the \matlab\
    function \texttt{nlhw} that uses the prediction error method to identify the
    linear block in the system (see~\cite{Ljung2009toolbox} for details). To get the best performance from this method, we
    equip it with an oracle that knows the true order of the LTI system generating the measurements.
\end{itemize}
We use two metrics to evaluate the performance of the estimators. We consider the
fitting score of the system impulse response
\begin{equation}
  FIT_{g,i} = 1 - \frac{\norm{g_i - \hat g_i}_2}{\norm{g_i - \bar g_i}_2} \,,
\end{equation}
where $g_i$ is the system generated at the $i$-th run of each experiment,
$\hat g_i$ its estimate and $\bar g_i$ its mean. A similar fitting score is
defined for the nonlinearity, namely
\begin{equation}
  FIT_{f,i} = 1 - \frac{\Vert f_i(u) - \hat f_i(u)\Vert_2}{\Vert f_i(u) -
  \overline{f_i(u)}\Vert_2} \,.
\end{equation}

Figure 2 shows one Monte Carlo realization with LTI system order equal to 10 and
SNR $= 10$, while Figure 3 shows the results of the outcomes of the 8
experiments. The box-plots compare the results of KB and NLHW for the considered
model orders. We can see that, for low model orders, the estimator NLHW equipped
with the true model order outperforms the proposed method. For higher model
orders, however, the proposed method KB-H gives substantially better performance
than NLHW\@. The reason is that KB-H is not affected by the increasing complexity (model
order) of the system, and the fitting score remains approximately constant. In
addition, we can notice that the estimation of the nonlinear block computed with
KB-H always provides a higher accuracy than NLHW\@.
\begin{figure}[!ht]
  \begin{center}
    \includegraphics[width=0.7\textwidth]{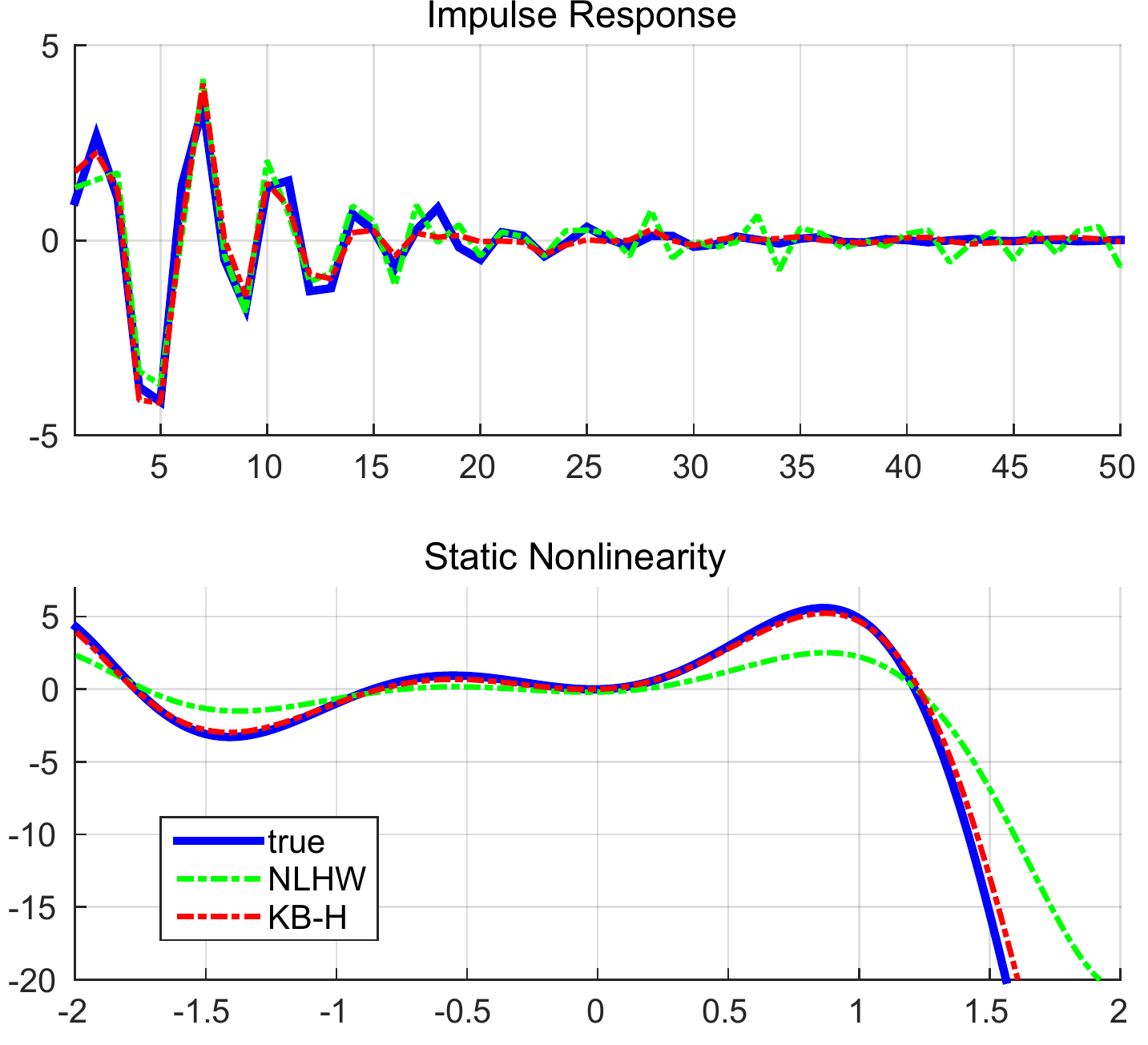}
    \end{center}
  \caption{Realizations of one Monte Carlo run with LTI system order 10 and SNR $= 10$.}\label{fig:realization}
\end{figure}
\begin{figure*}[!ht]
  \begin{center}
  \begin{tabular}{cc}
  \includegraphics[width=0.8\textwidth]{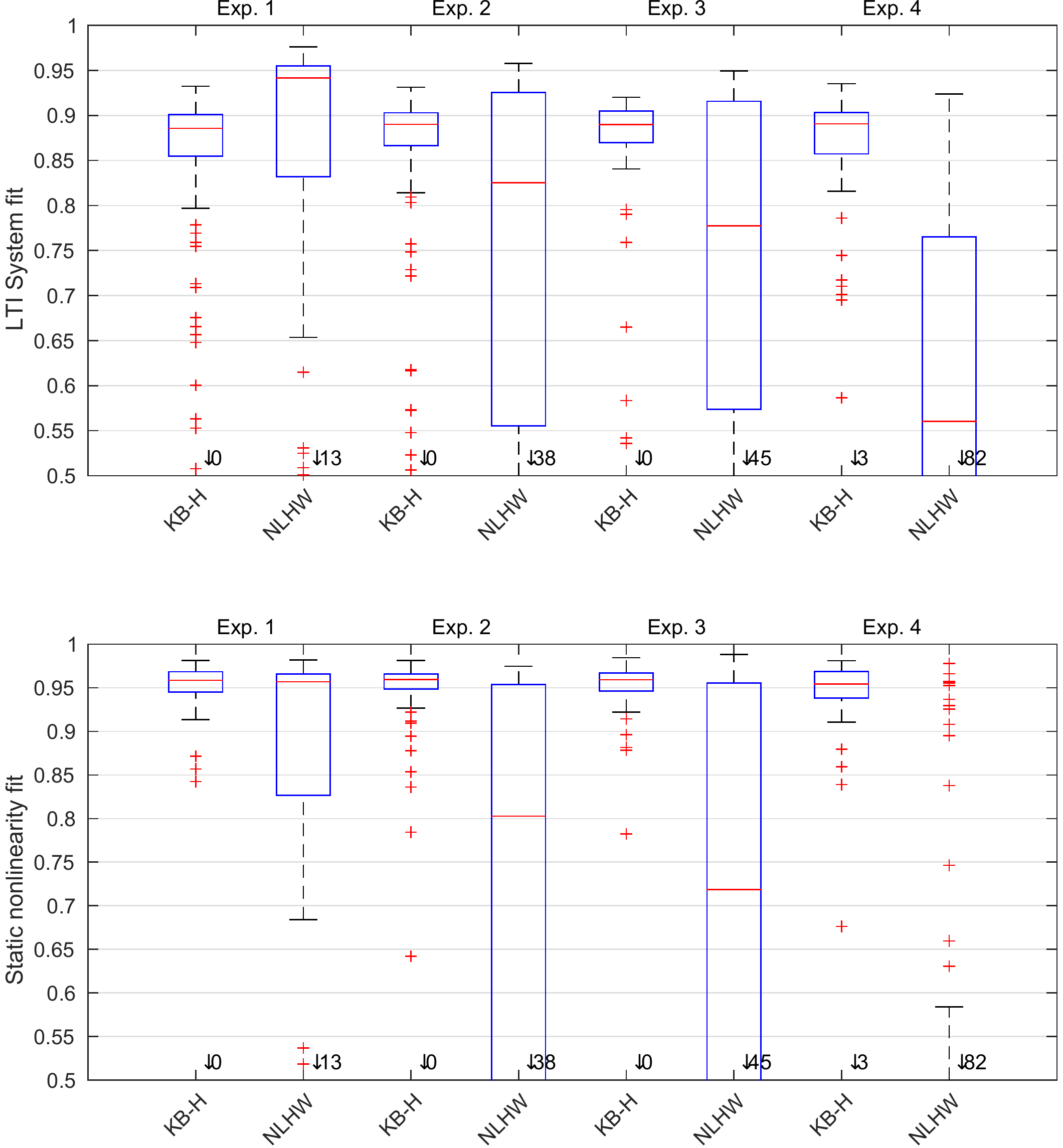}\\
  \includegraphics[width=0.8\textwidth]{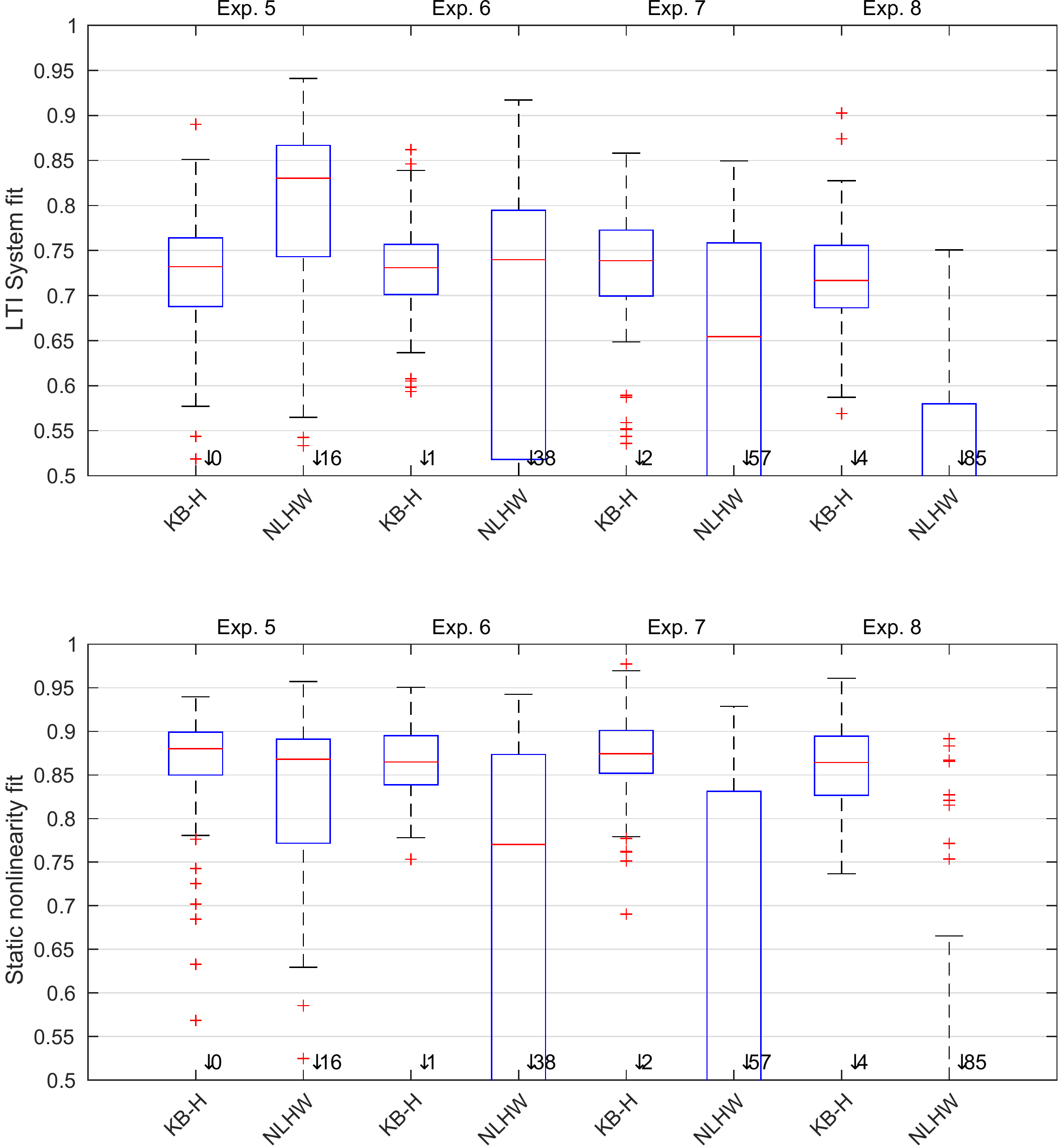}
  \end{tabular}
  \caption{Results of the 8 Monte Carlo experiments summarized in Table 1.}\label{fig:boxplot_10}
    \end{center}
\end{figure*}

\section{Conclusions}
In this work, we have proposed a novel kernel-based approach to the
identification of Hammerstein dynamic systems.  To model the impulse response we
have adopted a Gaussian regression approach and employed the stable spline
kernel. The identification of the input nonlinearity, together with the kernel
hyperparameter and the  noise variance, has been performed using an empirical
Bayes approach. The related marginal likelihood maximization has been carried
out resorting to the EM method. We have shown that this approach leads to an
iterative scheme consisting of a set of simple update rules, which allow for
fast computation. The effectiveness of the proposed method has been tested by
means of several numerical experiments. When compared with standard
state-of-the-art algorithms, the proposed method has shown a better fitting
capacity in both the nonlinear block and the LTI system impulse response.

We are currently working on the extension of the algorithm to a wider class of
system models. Furthermore, nonparametric descriptions of the nonlinear function
will be considered in order to obtain a completely parameter-free identification
method.

\appendix
\section{Proof of Theorem~\ref{th:main}}
The proof runs along the same arguments as the proof of Theorem 1 in~\cite{bottegal2015blind} and is included here for the sake of self-completeness.

Using the conditional probability formula,
\begin{equation}
  p(y,\,g;\,\theta) = p(y|g;\,\theta)p(g;\,\theta) \,.
\end{equation}
we can write the complete-data log-likelihood~\eqref{eq:complete_likelihood} as
\begin{equation}
  L(y,\,g;\,\theta) = \log p(y|g;\,\theta) + \log p(g;\,\theta)
\end{equation}
and so
\begin{align*}
    L(y,\,g;\,\theta) &=  -\frac{N}{2}\log \sigma^2 - \frac{1}{2\sigma^2}\norm{y - Wg}^2\allowdisplaybreaks[1]
      - \frac{1}{2}\log\det K_\beta - \frac{1}{2}g^T K_\beta^{-1}g \allowdisplaybreaks[1] \\
     & = -\frac{N}{2}\log \sigma^2 - \frac{1}{2\sigma^2}\left( y^T y +  g^T W^T Wg  -2y^T Wg    \right)\allowdisplaybreaks[1]  \\
     & \qquad\quad   - \frac{1}{2}\log\det K_\beta - \frac{1}{2}g^T K_\beta^{-1}g     \allowdisplaybreaks[1]  \,.
\end{align*}
We now proceed by taking the expectation of this expression with respect to the
random variable $g|y;\,\hat \theta^{(k)}$. We obtain the following components
\begin{align*}
  (a)&: \E\left[ -\frac{N}{2}\log\sigma^2 \right] \!=\!
    -\frac{N}{2}\log\sigma^2 \allowdisplaybreaks[1] \nonumber\\
  (b)&: \E\left[ -\frac{1}{2\sigma^2}y^T y \right] \!=\!
    -\frac{1}{2\sigma^2}y^T y \allowdisplaybreaks[1] \nonumber \\
  (c)&: \E\left[ -\frac{1}{2\sigma^2}g^T W^T Wg \right] = \nonumber \\
  & \qquad\qquad\qquad\trace\left[ -\frac{1}{2\sigma^2}W^T W(\hat P^{(k)}\! +\! \hat m_g^{(k)}\hat m_g^{(k)T}) \right] \allowdisplaybreaks[1] \nonumber\\
  (d)&: \E\left[ \frac{1}{\sigma^2}y^T Wg \right] \!=\!
    \frac{1}{\sigma^2}y^T W \hat m_g^{(k)} \allowdisplaybreaks[1] \nonumber\\
  (e)&: \E\left[ -\frac{1}{2}\log\det K_\beta \right] \!=\!
    -\frac{1}{2}\log\det K_\beta \allowdisplaybreaks[1] \nonumber\\
  (f)&: \E\left[ -\frac{1}{2}g^T K_{\beta}^{-1}g \right] =
    -\frac{1}{2}\trace \left[ K_\beta^{-1}( \hat P^{(k)} +  \hat m_g^{(k)}\hat m_g^{(k)T})  \right]
  \label{eq:E-step}
\end{align*}
It follows that $Q(\theta,\,\hat \theta^{(k)})$ is the summation of the elements
obtained above.  By inspecting the structure of $Q(\theta,\,\hat \theta^{(k)})$,
it can be seen that such a function splits in two independent terms, namely
\begin{equation}
  \mathcal{Q}(\theta,\,\hat\theta^{(k)}) = \mathcal{Q}_1(c,\,\sigma^2,\,\hat\theta^{(k)}) + \mathcal{Q}_\beta(\beta,\,\hat\theta^{(k)}) \,,
\end{equation}
where
\begin{equation}\label{eq:Q_1}
    \mathcal{Q}_1(c,\,\sigma^2,\,\hat\theta^{(k)}) = (a) +(b)+ (c)+ (d)
\end{equation}
is function of $c$ and $\sigma^2$, while
\begin{equation}\label{eq:Q_beta_appendix}
    \mathcal{Q}_\beta(\beta,\,\hat\theta^{(k)})  =  (e)+ (f)
\end{equation}
depends only on $\beta$ and corresponds to~\eqref{eq:Q_beta}. We now address the
optimization of~\eqref{eq:Q_1}. To this end we write
\begin{align}
  \mathcal{Q}_1(c,\sigma^2,\,\hat\theta^{(k)}) & = \frac{1}{\sigma^2}\mathcal{Q}_c(c,\,\hat\theta^{(k)}) + \mathcal{Q}_{\sigma^2}(\sigma^2,\,\hat\theta^{(k)}) \\
   & = \frac{1}{\sigma^2} \Big(\trace\left[-\frac{1}{2}W^T W(\hat P^{(k)} + \hat m_g^{(k)}\hat m_g^{k T}) \right] \nonumber \\
   & + y^T W \hat m_g^{(k)} \Big)  -\frac{N}{2}\log\sigma^2 -\frac{1}{2\sigma^2}y^T y  \nonumber
  \label{eq:Q_1_split}
\end{align}

This means that the optimization of $\mathcal{Q}_1$ can be carried out first
with respect to $c$, optimizing only the term $\mathcal{Q}_c$, which is
independent of $\sigma^2$ and can be written in a quadratic form
\begin{equation}\label{eq:quadratic_cost}
  \mathcal{Q}_c(c,\,\hat\theta^{(k)}) =  -\frac{1}{2}c^T\hat A^{(k)} c + \hat b^{(k)T}c \,.
\end{equation}
To this end, first note that, for all $v_1\in \mathbb{R}^n$, $v_2 \in
\mathbb{R}^m$:
\begin{equation}
  \T_m(v_1)v_2 = \T_n(v_2)v_1 \,.
\end{equation}
Recalling~\eqref{eq:devec}, we can write
\begin{equation*}
  \begin{split}
    & -\frac{1}{2}\trace \left[W^T W( \hat P^{(k)}\!\! +\!\!  \hat m_g^{(k)}\hat m_g^{(k)T})  \right] \\
    & =  -\frac{1}{2}{\vect(W)}^T\! (( \hat P^{(k)} \!\!+\!\!  \hat m_g^{(k)}\hat m_g^{(k)T}) \kron I_{N} ) \vect(W)\\
    &= -\frac{1}{2}w^T\R^T\left(( \hat P^{(k)} +  \hat m_g^{(k)}\hat m_g^{(k)T}) \kron
      I_{N}\right)\R w\\
      &= -\frac{1}{2}c^T {F(u)}^T\R^T\left(( \hat P^{(k)} +  \hat m_g^{(k)}\hat m_g^{(k)T}) \kron I_{N} \right)\R F(u)c \,,
  \end{split}
\end{equation*}
where the matrix in the middle corresponds to $\hat A^{(k)}$ defined in~\eqref{eq:A_b}. For the linear term we find
\begin{equation}
y^T W \hat g^{(k)}  =  y^T\T_N(\hat m_g^{(k)})w = y^T\T_N(\hat m_g^{(k)})F(u)c \,,
\end{equation}
so that the term $\hat b^{(k)T}$ in~\eqref{eq:A_b} is retrieved and the
maximizer $\hat c^{(k+1)}$ is as in~\eqref{eq:new_c}.  Plugging back
$\hat c^{(k+1)}$ into~\eqref{eq:Q_1} and maximizing with respect to $\sigma^2$
we easily find $\hat \sigma^{2,(k+1)}$ corresponding to~\eqref{eq:new_sigma}.
This concludes the proof.

\printbibliography{}
 \end{document}